\documentclass[12pt,a4paper]{article}
\usepackage{authblk}
\usepackage{graphicx}

\begin{document}

\title{An Evaluation and Enhancement of Seredynski-Bouvry CA-based Encryption Scheme}
\author[1]{Hossein Arabnezhad\thanks{ arabnezhad@aut.ac.ir}}
\author[2]{Babak Sadeghiyan\thanks{basadegh@aut.ac.ir}}
\affil[1,2]{Department of Computer Engineering, Amirkabir University of Technology}
\date{}
\maketitle

\pagestyle{plain}
\begin{abstract}
In this paper, we study a block cipher based on cellular automata, proposed by Seredynski and Bouvry in \cite{semabo04} against \emph{plain-text avalanche criteria} and \emph{randomness} tests. Our experiments shows that Seredynski-Bouvry encryption scheme does not pass some NIST statistical tests by neighborhood radius less than three. It also showed that if the CA rule is selected carelessly, it weaken the security of scheme. Therefor, the selection of CA-rule as part of key can not be left to the user. Hence, cryptographic properties such as balancedness and non-linearity should be considered in the selection of CA-rules. This approach is more compliant with Kerckhoffs principle. So security should depend just on security of final data. We also improve Seredynski-Bouvry encryption scheme to satisfy strict avalanche criteria and NIST statistical test suite in about half number of iterations comparing to original scheme. This improvement is achieved by change in the definition of neighborhood.
\end{abstract}

\section{Introduction}

Nowadays private key cryptosystems have an important role in security of commercial communications.
In this type of cryptosystems, sender and receiver share a secret key privately. \emph{Advanced Encryption Standard}(AES) is an example of private key block ciphers standardized for confidentiality purposes.
Block ciphers are a subclass of private key cryptosystems, in which data is divided into blocks and each block encrypted individually. 
Design of these cipher is based on two key properties: confusion and diffusion. 
One of main features of cellular automata is to emerge complex behavior by application of simple rules, frequently, on array of cells.
 This feature has conformance to block cipher design objectives. 
In design of block ciphers we try to build a complex relation between plain-text and cipher-text by application of a simpler round function, iteratively. 
So, cellular automata can be a natural solution to design of block cipher.

Many CA-based cryptography schemes proposed in the literature, including public key, stream ciphers \cite{seredynski2004cellular} and block ciphers \cite{semabo04,sensha02cac}. 
There are also schemes that build up a cipher based on CA-based components such as s-box or linear transformation \cite{ghamey99caapps}.

One of these scheme is proposed by seredynski and bouvry in \cite{semabo04}. Their proposed scheme is based on reversible cellular automata. 
They studied their scheme only against strict avalanche criteria. 
In this article we verify the results presented in \cite{semabo04}.
A secure block cipher also can be used as a pseudo random number generator. We study this scheme against NIST statistical test suite for testing randomness of generated bitstreams. 
We improve scheme to satisfy mentioned properties in half number of iterations comparing to original scheme.

In section \ref{sec:prelim} we present some preliminaries about elementary CA and reversible CA. In section \ref{sec:serboscheme}, Seredynski-Bouvry encryption scheme is introduced. Our study of Seredynski-Bouvry scheme is presented in \ref{sec:anal}.
\section{Preliminaries \label{sec:prelim}}
\subsection{Elementary Cellular Automata}
Elementary CA(ECA) is an array of cells, in which, each cell is assigned a value from a set of state alphabet. 
All cells are updated synchronously by application of a rule $R$ in discrete times.
An ECA can be explicitly expressed by its size, the rules to be applied, initial state and formation of neighborhood. 
Size of CA defines number of cells contained in CA. When CA size is finite, usually, it is considered as a ring. 
Each application of rule to all cell in CA at time $t$ is called an iteration. 
Before first iteration cells may be assigned values which is called initial state. 
Next state of each cell $S_i^{t+1}$ at time  $t$ can be obtained by :
\begin{equation}
\label{eq:ecarule}
S_i^{t+1} = R(S_{i-r}^t,...,S_{i-1}^t,S_{i}^t,S_{i+1}^t,...,S_{i+r}^t).
\end{equation}
In Eq. \ref{eq:ecarule}, $r$ defines neighborhood radius and $S_i^t$ shows the state of $i$-th cell. Number of possible rules is $2^k$ in which $k= 2^{2r+1}$. Proposed CA in \cite{semabo04} is a one-dimensinal uniform CA with binary alphabet $\{0,1\}$. Neighborhood radius is either 2 or 3, for size 32 and 64 cells.

\subsection{Reversible Cellular Automata}
\begin{figure}[h]
\centering
\includegraphics[scale=1]{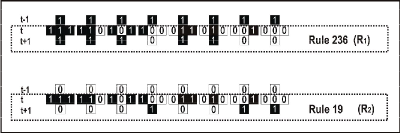}
\caption{reversible rule 236/19 \cite{semabo04}}
\label{fig:rul}
\end{figure}

By application of rules on each cell $s_i$ in configuration $q_t$, we get a new configuration $q_{t+1}$. 
We can view this transformation as a global transition function which map configuration $q_t$ to $q_{t+1}$. 
A CA is reversible if only if its global transition function is one-to-one. 
Study of elementary CA showed only few rules have this property. 
For example, considering all 256 rule with neighborhood radius 1, only six rules are reversible. 
Hence, a new class of reversible introduced in \cite{wolfram2002new}. 
In addition to current state of cell, rules in this reversible CA are depending on state of cell in the previous iteration. 
Eq. \ref{eq:rcarule} defines how the rule applies to a cell and its neighbors.
\begin{equation}
\label{eq:rcarule}
S_i^{t+1} = R(S_{i-r}^t,...,S_{i-1}^t,S_{i}^t,S_{i}^{t-1},S_{i+1}^t,...,S_{i+r}^t)
\end{equation}
In Eq. \ref{eq:rcarule} $S_i^{t+1}$ depends on state of cell in iterations $t$ and $t-1$. 
These rules can be easily derived from ECA rules.
The rule in RCA is comprised of two elemetary rules. 
One is applied when the cell state is 1, otherwise second rule is applied. 
Figure \ref{fig:rul} shows an example of such rules.
If the first rule is known $R_1$, the second rule can be easily derived by following equation: 
\begin{equation}
R_2 = 2^d - R_1 -1.
\label{eq:rul}
\end{equation}
in which, $d = 2^{2r+1}$ and $r$ is the neighborhood radius.
Considering that each of these rules are depending on state in the previous iteration, the initial configuration of CA is comprised of two initial configuration $q_0$ and $q_1$. These rules are also used to proceed in backward or forward direction.

\section{Seredynski-Bouvry encryption scheme\label{sec:serboscheme}}
\begin{figure}[h]
\centering
\includegraphics[width=12cm]{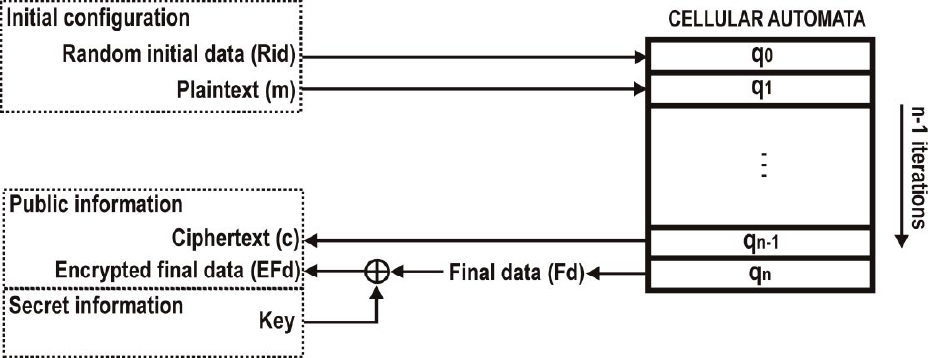}
\caption{Encrypting a block of data by Seredynski-Bouvry scheme\cite{semabo04}}
\label{fig:enc}
\end{figure}

\begin{figure}
\centering
\includegraphics[width=12cm]{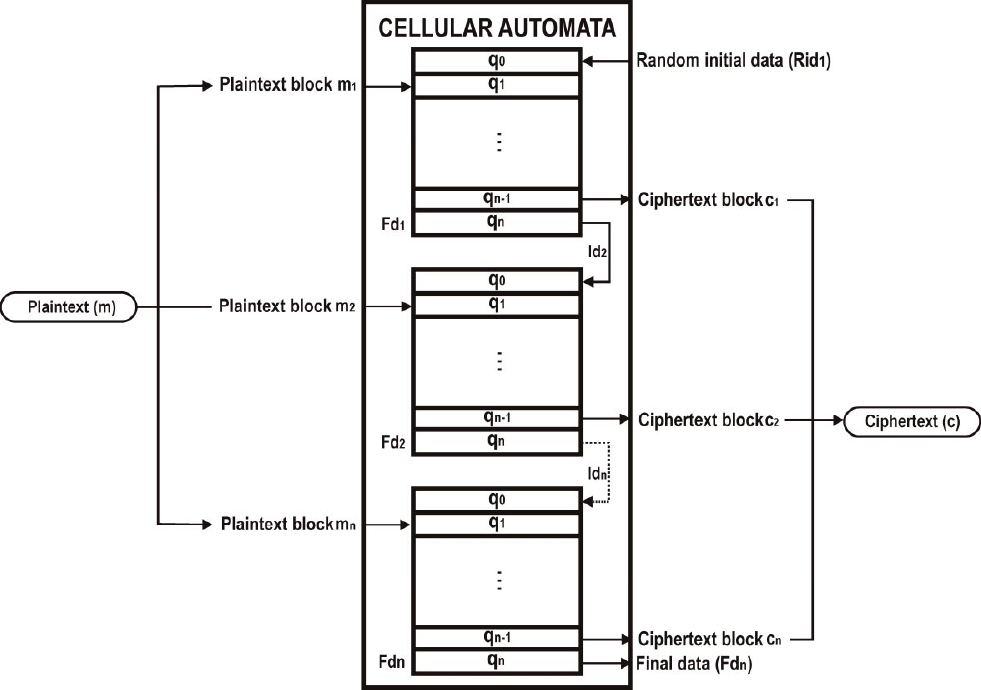}
\caption{Encryption process of multi-block data \cite{semabo04}}
\label{fig:mulblk}
\end{figure}

\begin{figure}
\centering
\includegraphics[width=12cm]{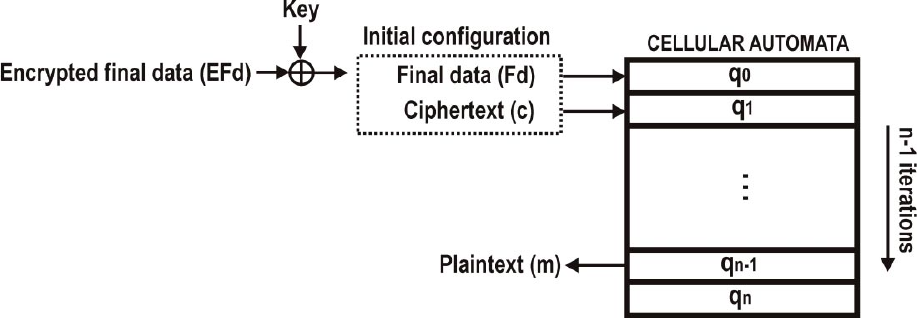}
\caption{Decrypting a block of data by Seredynski-Bouvry scheme\cite{semabo04}}
\label{fig:dec}
\end{figure}

In this section we introduce Seredynski-Bouvry CA-based block cipher \cite{semabo04}. In this scheme, plain-text is encoded as a part of initial configuration, $q_1$. $q_0$ part of initial configuration is filled by random data. Encryption process is done through iterating in forward direction.  This process illustrated in figure \ref{fig:enc}.
After $n-1$ iterations, configuration $q_{n-1}$ contains cipher-text. In this scheme, $q_n$ must be delivered to other party securely, by encryption or through a secure channel.
If plain-text length is more than one block, it is divided into $n$ blocks, and each block is encrypted separately. $q_0$ of first block is filled by random data. $q_0$ configuration of other blocks are filled by final data of previous block. Final data of last block, $q_n$, must delivered to other party securely. Figure \ref{fig:mulblk} illustrates the encryption process for multi-block data.
Decryption process is depicted in figure \ref{fig:dec}. 
In decryption process initial configuration is composed of cipher-text and final data. After that $q_0$ and $q_1$ are filled by final data and cipher-text respectively and CA will be iterated $n-1$ times to obtain the plain-text.

\begin{figure}[h]
\centering
\includegraphics[width=12cm]{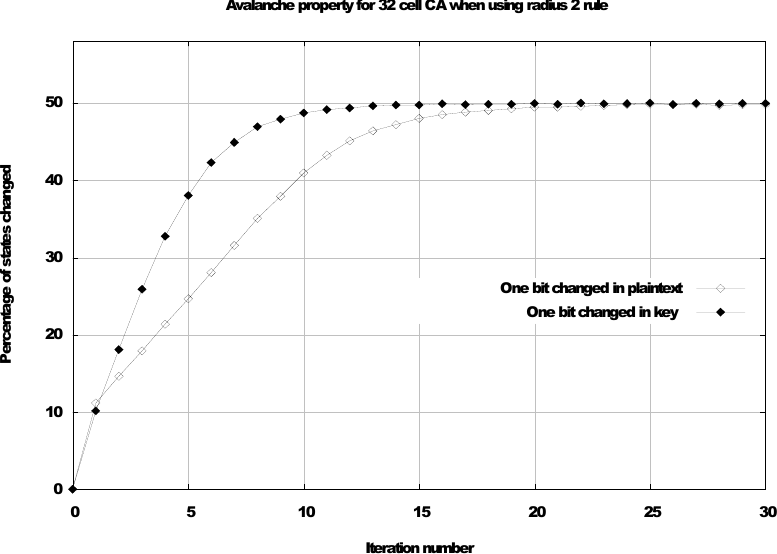}
\caption{Avalanche property for 32 cell CA and radius 2 rule\cite{semabo04}}
\label{fig:refplot}
\end{figure}

\begin{figure}[h]
\centering
\includegraphics[width=12cm]{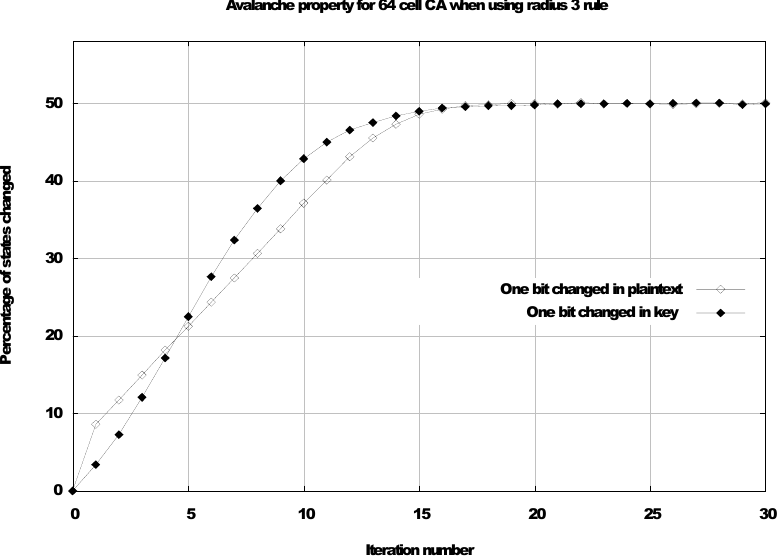}
\caption{Avalanche property for 64 cell CA and radius 3 rule\cite{semabo04}}
\label{fig:refplot2}
\end{figure}

So called avalanche property of a cipher, is that even a single change in plain-text or key must lead to significant changes in cipher-text. Strict avalanche criteria states that by any change in plain-text or key, each bit of cipher-text must be flipped with probability 0.5. In \cite{semabo04}, Seredynski-Bouvry scheme is studied against this criteria for CA with size 32 and 64 cells. Neighborhood radius is taken 2 and 3 for CA size 32 and 64, respectively. Figures \ref{fig:refplot} and \ref{fig:refplot2} show relation of number of iteration and behavior of the proposed cipher against strict avalanche criteria for CA with size 32 and 64, respectively. As stated in \cite{semabo04}, initial data, rules are selected randomly.

\section{Our study of Seredynski-Bouvry scheme \label{sec:anal}}
We implemented Seredynski-Bouvry encryption scheme using C++ in Ubuntu. Just as \cite{semabo04}, we also select CA size to 32 and 64 cells.

\subsection{Strict Avalanche Criteria}
\begin{figure}[h]
\centering
\includegraphics[width=12cm]{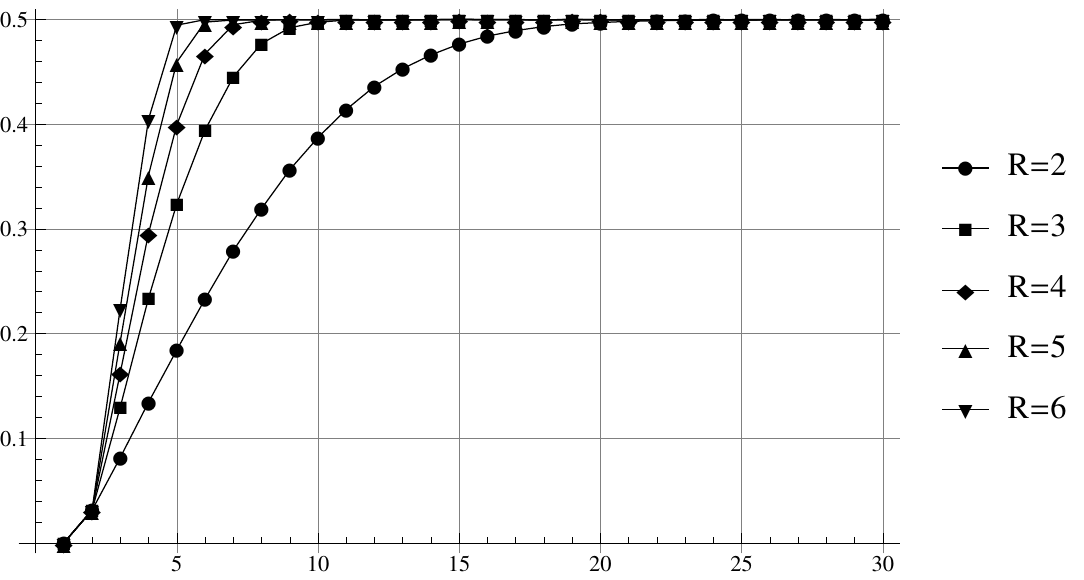}
\caption{strict avalanche criteria for CA with 32 cells and radius $r=2$ to  $r=6$}
\label{fig:plot}
\end{figure}

\begin{figure}[h]
\centering
\includegraphics[width=12cm]{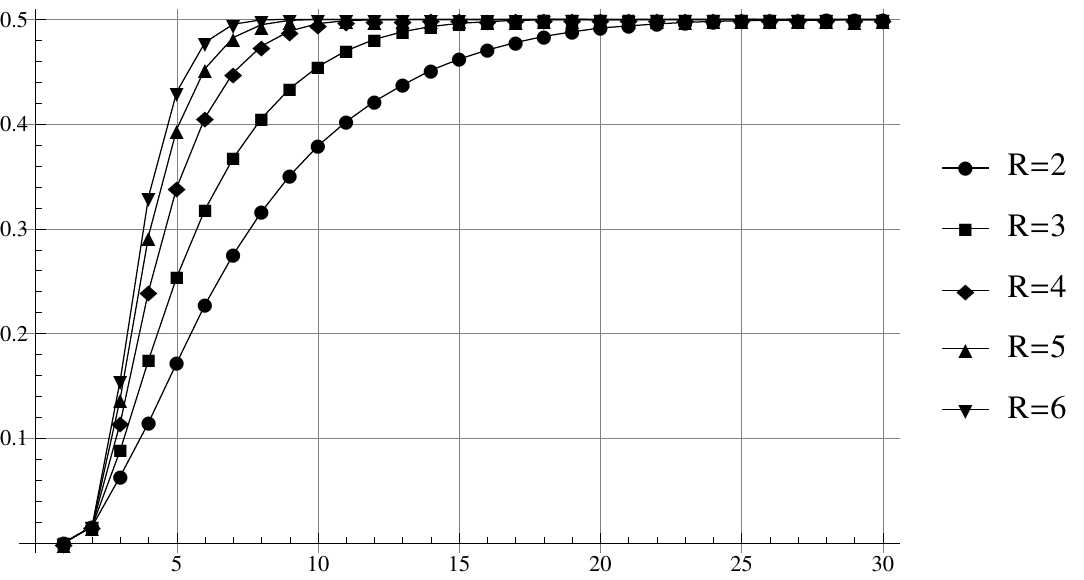}
\caption{strict avalanche criteria for CA with 64 cells and radius $r=2$ to  $r=6$}
\label{fig:plot64}
\end{figure}

For our experiments, we generate a random rule and obtain its complement by \ref{eq:rul}.
This experience is done over 10000 random plain-texts and initial data. In our test, the rule is generated randomly and it is fixed for each neighborhood radius.
Figures \ref{fig:plot} and \ref{fig:plot64}  illustrate results for CA with sizes 32 and 64 cells. 

These result are compatible with results presented in \cite{semabo04}. 
As illustrated in figures \ref{fig:plot} and \ref{fig:plot64}, greater neighborhood radius leads to satisfaction of SAC in smaller number of iterations. 
So, we can establish an trade-off of between performance and security.  
\subsection{Statistical testing of randomness}
Block cipher are also can be used as a pseudo random number generator for various cryptographic application.
So sequence of bits generated by the underlying block cipher, must be random. 
NIST statistical test suite, is a collection statistical test to verify whether a sequence of bits is whether random. 
NIST statistical tests is comprised of 15 different tests. 

In table \ref{tab:nist}, we bring results of some of tests, which may not be passed for radius two, for different neighborhood radius.
As it showed in table \ref{tab:nist}, CA with neighborhood size two can not pass statistical tests. So the size of neighborhood radius should be greater than 3.
\begin{table}
\centering
\caption{NIST Statistical Tests}
\label{tab:nist}
\begin{tabular}{|c|c|c|c|c|c|}
\hline
Statistical Test /neighborhood radius & $r=2$ & $r=3$ & $r=4$ & $r=5$ & $r=6$ \\
\hline
 Frequency & $ - $ & $  + $ & $ + $ & $ + $ & $ + $ \\
\hline
 Frequency in block& $ - $ & $ + $ & $ + $ & $ + $ & $ + $ \\
\hline
Run & $ + $ & $ + $ & $ + $ & $ + $ & $ + $ \\
\hline
Binary Matrix Rank & $ - $ & $ + $ & $ + $ & $ + $ & $ + $ \\
\hline
Fast Fourier Transform & $ - $ & $ + $ & $ + $ & $ + $ & $ + $ \\
\hline
Universal Maurer Test& $ - $ & $ + $ & $ + $ & $ + $ & $ + $ \\
\hline
\end{tabular}
\end{table}

\subsection{Effect of rules}
\begin{figure}[h]
\centering
\includegraphics[width=12cm]{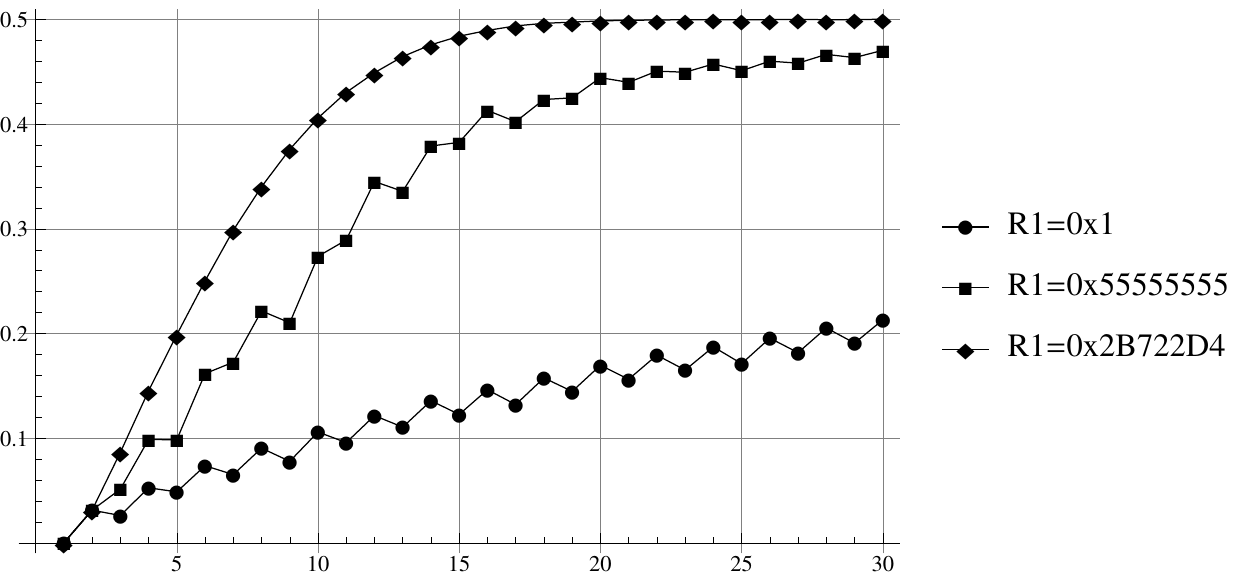}
\caption{SAC for CA with three rules $0x1$, $0x55555555$  and $0x2B722D4$. }
\label{fig:plotrul}
\end{figure}

In this scheme, rules are part of key and naturally selected randomly. Selected rules have direct effect on cryptographic properties of the cipher. We study the effect rule to SAC for three rules: 0x1, 0x55555555, 0x2B722D4. In \ref{fig:plotrul} illustartes effect of these rule on satisfying SAC, for a CA with size 32 and neighborhood radius two. As it showed, selection of rule have direct effect on satisfying cryptographic properties, so rules should be selected carefully and cannot be selected as a part of key just by a random process.

\begin{table}
\centering
\caption{boolean function equivalent to three rules $0x1$, $0x55555555$  and $0x2B722D4$.}
\label{tab:bool}
\begin{tabular}{|c|c|}
\hline
Rule & Equivalent Boolean Function \\
\hline
 $0x1$ & $\neg {a}\land \neg {b}\land \neg {c}\land \neg {d}\land \neg {e}$\\
\hline
 $0x55555555$ & $\neg e$\\
\hline
 $0x2B722D4$ & $\begin{array} {c} (a\land \neg b\land \neg c\land \neg d)\oplus (b\land \neg c\land \neg d\land e)\oplus (\neg b\land c\land d\land e)\oplus  \\ (\neg b\land c\land \neg d\land \neg e)\oplus (\neg b\land \neg c\land d\land \neg e)\oplus (a\land \neg b\land c\land \neg d\land e)\oplus  \\ (\neg a\land b\land c\land \neg d\land e)\oplus (\neg a\land \neg b\land c\land d\land \neg e)\end{array}$ \\
\hline
\end{tabular}
\end{table}
\section{Improvement of Seredynski-Bouvry Scheme}
\begin{figure}[h]
\centering
\includegraphics[width=12cm]{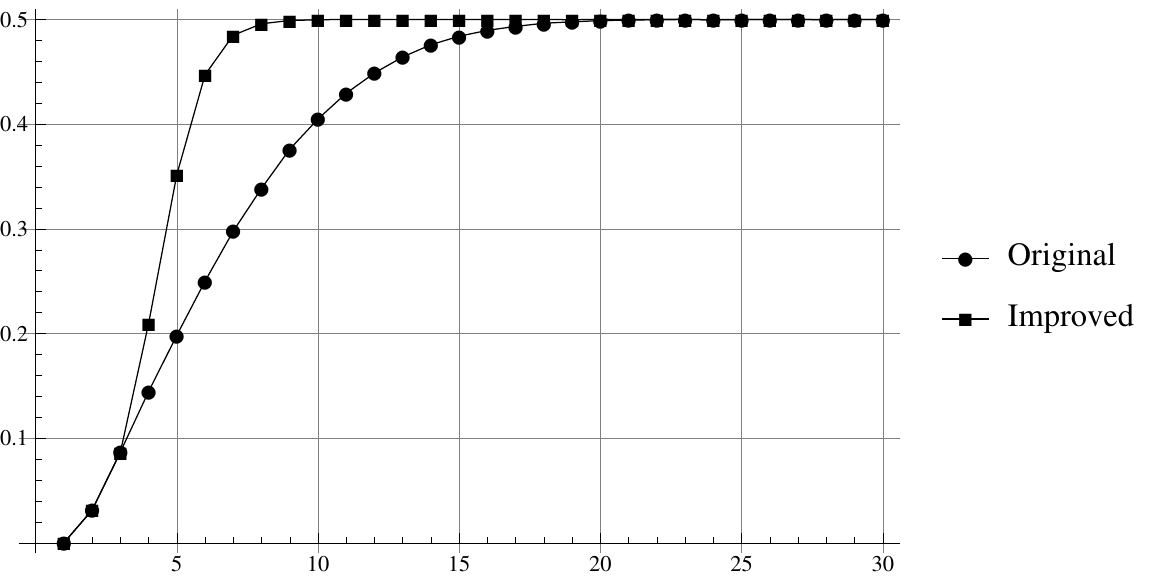}
\caption{comparison of effect of neighborhood definition with \cite{semabo04} and ours}
\label{fig:plotimpl}
\end{figure}

The neighborhood defined in \cite{semabo04} is a kind of normal neighborhood, that is, the propagation of changes in state of cells of array to other cells is not fast enough. We can improve propagation by changing the neighborhood definition. The new neighborhood defined in Eq. \ref{eq:imp}. 
\begin{equation}
\label{eq:imp}
S_{5i}^{t+1} = R(S_{i-r}^t,...,S_{i-1}^t,S_{i}^t,S_{i}^{t-1},S_{i+1}^t,...,S_{i+r}^t)
\end{equation}
In Eq. \ref{eq:imp}, next state of $S_{5i}$ is obtained through application of rule on cells $i-r$ through $i+r$.
In figure \ref{fig:plotimpl} we compare two neighbourhood definitions. As we see our definition of neighbourhood, leads to satisfaction of SAC in smaller number of iterations (about half).

\section{Conclusion}
In this paper we studied Seredynski and Bouvry encryption scheme \cite{semabo04} against strict avalanche criteria and NIST randomness tests. 
We also studied effect rule in satisfying aforementioned criteria. It showed that randomly selected rules may weaken security of scheme. So, rules must be selected with respect to cryptographic properties such as balancedness and completeness. We changed neighbourhood definition, such that 
strict avalanche criteria can be satisfied in half number of iterations comparing to original scheme.
\bibliographystyle{ieeetr}
\bibliography{article}
\end{document}